\documentclass[aps,prb,superscriptaddress,draft,showpacs,intlimits,amsmath,amssymb,floats,floatfix,twocolumn]{revtex4}
\usepackage[final]{graphicx}
\usepackage{epsfig}
\begin{document}
\title{Electron-boson glue function derived from electronic Raman scattering}
\author{B. Muschler}
\affiliation{Walther Meissner Institut, Bayerische Akademie der
Wissenschaften, 85748 Garching, Germany}
\author{W. Prestel}
\affiliation{Walther Meissner Institut, Bayerische Akademie der
Wissenschaften, 85748 Garching, Germany}
\author{E. Schachinger}
\email{schachinger@itp.tu-graz.ac.at}
\affiliation{Institute of Theoretical and Computational Physics,
Graz University of Technology, A-8010 Graz, Austria}
\author{J. P.~Carbotte}
\affiliation{Department of Physics and Astronomy, McMaster University, 
Hamilton, Ontario N1G 2W1, Canada}
\affiliation{The Canadian Institute for Advanced Research, Toronto,
Ontario M5G 1Z8, Canada}
\author{R. Hackl}
\affiliation{Walther Meissner Institut, Bayerische Akademie der
Wissenschaften, 85748 Garching, Germany}
\author{Shimpei~Ono}
\affiliation{Central Research Institute of Electric Power Industry,
              Yokosuka, Kanagawa 240-0196, Japan}
\author{Yoichi Ando}
\affiliation{Institute of Scientific and Industrial Research,
              Osaka University, Ibaraki, Osaka 567-0047, Japan}

\begin{abstract}
Raman scattering cross sections depend on photon polarization. In the
cuprates
nodal and antinodal directions are weighted more strongly in $B_{2g}$ and
$B_{1g}$ symmetry, respectively. On the other hand in angle-resolved
photoemission spectroscopy (ARPES), electronic properties
are measured along well-defined directions in momentum space rather
than their weighted averages.
In contrast, the optical conductivity involves a momentum
average over the entire Brillouin zone. Newly measured
Raman response data on high-quality Bi$_2$Sr$_2$CaCu$_2$O$_{8+\delta}$
single crystals up to high energies have been inverted using a
modified maximum entropy inversion
technique to extract from $B_{1g}$ and $B_{2g}$ Raman data corresponding
electron-boson spectral densities (glue) are compared to
the results obtained with
known ARPES and optical inversions. We find that the $B_{2g}$ spectrum
agrees qualitatively with nodal direction ARPES while the $B_{1g}$
looks more like the optical spectrum. A large peak around $30 - 40\,$meV
in $B_{1g}$, much less prominent in $B_{2g}$, is taken as support for the
importance of $(\pi,\pi)$ scattering at this frequency.
\end{abstract}

\maketitle

Boson structures seen in tunneling and to a lesser extent in optics in
conventional superconductors have given us detailed information about
the electron-phonon interaction in these materials. While similar
structures have been identified in the cuprates in tunneling,
point contact junctions \cite{zas06} as well as
scanning tunneling spectroscopy (STS) \cite{lee06,pasupathy08},
in angle-resolved photo emission (ARPES)
\cite{Lanzara:2001,cuk04,zhou05,meevasana06,valla07,kordyuk06,schach08a},
and, particularly, in optics \cite{schach00,hwang07a,schach03,
heumen09}, some of the details associated with the recovered electron-boson
spectral density remain controversial, particularly the nature of the
bosons involved. Some investigators stress the role of phonons
\cite{Lanzara:2001,zas06,lee06,pasupathy08,cuk04,zhou05,meevasana06} while
others favour spin fluctuations \cite{valla07,
kordyuk06,schach08a,schach00,hwang07a,schach03,
heumen09}. Of course, both mechanisms are
expected to contribute to the effective electron-boson interaction
so that the debate really centers on which might be dominant \cite{heumen09,
schach09,iwasawa08}. Recent ARPES data have found a 6\% softening of the
boson `kink' in the nodal direction renormalized electronic dispersion
curves for Bi$_2$Sr$_2$CaCu$_2$O$_{8+\delta}$ (Bi2212) compounds on
substitution of $^{16}$O by $^{18}$O \cite{schach09,iwasawa08}.
Based on an examination of the
detailed shape of the electron-boson spectral function, $I^2\chi(\omega)$,
(glue function) obtained from a maximum entropy inversion \cite{schach06}
of data on a similar sample \cite{schach08a}
of Bi2212, Schachinger {\it et al.} \cite{schach09} have argued that
the isotope substitution data can be understood if a peak seen in
the electron-boson spectral density
$I^2\chi(\omega)$ around $65\,$meV which contains about 10\% of
the total area is assigned to an oxygen phonon mode. This leaves
90\% of the effective spectrum which extends to $400\,$meV, much larger
than any phonon energy, to possibly come from spin fluctuations.

So far $I^2\chi(\omega)$ has been
recovered from tunneling \cite{zas06,zas01},
nodal direction ARPES \cite{zhou05,meevasana06,valla07,schach08a},
and optical data in the cuprates \cite{schach00,hwang07a,
yang09,carb05,schach03,heumen09}.
In principle, $I^2\chi(\omega)$ is anisotropic and it
will be different for each momentum direction. Also quasiparticle and
optical spectral densities will not be the same. While both involve
the same bosons different weighting electronic factors apply. This is also
true for Raman for which different vertices apply for different photon
polarization \cite{Devereaux:1994,branch00,branch95,jiang96,opel00,
Devereaux:2007}. Nevertheless, it is important to obtain the corresponding
electron-boson spectral density associated with $B_{1g}$ and $B_{2g}$ Raman
data and to understand how these might differ from those obtained from
ARPES and from optics and to establish points of
consistency between these various spectra.
Based on a memory function approach to the Raman cross section \cite{opel00}
one can extract a corresponding Raman scattering rate which depends on
polarization. In this work we apply a maximum entropy inversion
technique to extract from such data a spectral density in analogy to what
is done for the infrared conductivity.

The analysis starts from new experimental results from freshly
prepared high-quality single crystals of Bi$_2$Sr$_2$CaCu$_2$O$_{8+\delta}$.
The crystals were grown in a mirror furnace using the traveling-solvent
floating-zone (TSFZ) technique and post-annealed at $870\,{^\circ}$C
to arrive at optimal doping ($p=0.16$ holes/CuO$_2$) with a $T_c$ of 94.5\,K
and a transition width below 1\,K. The quality of the samples is crucial since
defects lead to a strongly enhanced cross section at high
energies \cite{Muschler:2010}.

The basis of the analysis is the Raman response function
$R_{\mu}\chi''_{\mu}(T,\omega)$, where $\mu$ represents the
scattering symmetry. Here we focus on $\mu = B_{1g}$ and $B_{2g}$. The
response is derived from the measured cross section $\sigma_\mu$ as
$R_\mu\chi''_\mu = (\omega_I/\omega_S)[1+n(\omega,T)]^{-1}d^2\sigma_\mu/%
(d\Omega\, d\omega_S)$ \cite{Devereaux:2007}. Here,
$\omega_{I,(S)}$ is the frequency of the incident (scattered) light.
$\Omega$ is the solid angle of acceptance of the collection optics and
$n(\omega,T) = \{\exp[\hbar\omega/(k_B T)-1]\}^{-1}$ is the Bose factor.
$\sigma_\mu$ is corrected for the sensitivity of the instrument. Since
the optical constants vary only little for visible light, interface
effects can be absorbed in the constant $R_\mu$ the magnitude of which
is irrelevant here. For revealing the pure symmetry components of the
response in a reliable fashion we measured spectra at all six main
polarizations of the $ab$-plane. This allows us to check the internal
consistency of the procedure. Thus, the $B_{1g}$ response, for example,
reads in a short hand notation
\begin{equation}
\label{req:1}
B_{1g} = \frac{1}{3}\left[xx +x^\prime y^\prime+RL-\frac{1}{2}
 (RR+xy+x^\prime x^\prime)\right],
\end{equation}
where $x=[100]$, $x^\prime = [110]$, $R=[100]+i[010]$, etc.
The typical number of photon counts per point is between
1000 and 10000 which results in a relative statistical error in the
range of one to three percent. For the experiments shown here the
read-out noise of the CCD detector is negligable.

Opel {\it et al.} \cite{opel00} show how a Raman scattering
rate, $\Gamma_{\mu}(T,\omega)$, can be extracted quantitatively from
$R_{\mu}\chi''_{\mu}(T,\omega)$. At a given symmetry
the sensitivity in momentum space is uniquely defined, and the main
contributions in $B_{1g}$ and $B_{2g}$ symmetry come from the region
close to ($\pi,0$) (and equivalent points of the Brillouin zone) and from
the center of the quadrant, respectively. The derived scattering
rates $\Gamma_{\mu}(T,\omega)$ are analogous
to the optical scattering rate, $\tau^{-1}_{\rm opt}(T,\omega)$, obtained
from infrared measurements \cite{schach00} but correspond to different
parts of the BZ dictated by symmetry $\mu$ \cite{Devereaux:2007}.
In the normal state and within a Kubo formalism  we can show that
the Raman spectrum \cite{jiang96} and the optical
response \cite{nicol91,mars96,carb95,schach97,jiang96a}
is related to the appropriate electron-boson
spectral density, $I^2_{\mu}\chi(\nu)$, to a good approximation
through the equation \cite{mars98a,shulga91,sharapov05}
\begin{equation}
\label{eq:1}
 \Gamma_{\mu}(T,\omega) - \tau^{-1}_{\mu,imp}=
  \int\limits_0^\infty\!d\nu\,
  K(\omega,\nu;T) I^2_{\mu}\chi(\nu),
\end{equation}
where $\tau^{-1}_{\mu,imp}$ is a Raman impurity scattering rate and
\begin{eqnarray}
\label{eq:2}
 K(\omega,\nu;T) &=& \frac{\pi}{\omega}\left[ 2\omega\textrm{coth}\left(
  \frac{\nu}{2T}\right)-(\omega+\nu)\textrm{coth}\left(\frac{\omega+
  \nu}{2T}\right)\right.\nonumber\\
  &&\left.+(\omega-\nu)\textrm{coth}\left(\frac{\omega-\nu}{2T}
  \right)\right],
\end{eqnarray}
where $T$ denotes the temperature. For the conductivity which
involves an average
over all momentum directions $\theta$ the appropriate electron-boson
spectral density is
$I^2_\mu\chi(\nu) = \langle I^2_\mu\chi(\nu,\theta)\rangle_\theta$
with $\langle\cdots\rangle_\theta$ the average over the directions $\theta$.
For the Raman case with symmetry $\mu$ there is an additional
weighting of $\cos^2(2\theta)$ for $B_{1g}$ and
$\sin^2(2\theta)$
for $B_{2g}$. Here, $I^2_\mu\chi(\nu,\theta)$ is the electron-boson spectral
density associated with momentum direction $\theta$.
There is another difference between quasiparticle and transport
quantities associated with vertex corrections. These are expected to
mainly change the magnitude of the distribution functions with shape
changes secondary.

\begin{figure*}[tp]
  \begin{center}
  \includegraphics[width=10cm]{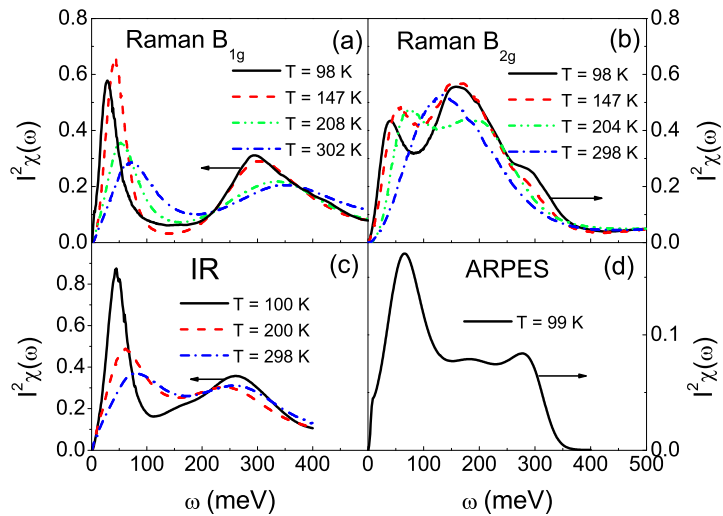}
  \end{center}
  \caption{(Colour on-line) The electron-boson spectral density
(dimensionless) as a function of energy $\omega$ in meV from $B_{1g}$
(a) and $B_{2g}$ (b) Raman data. We show in (c)
results \cite{schach06} obtained from optical data
\cite{tu02} and in (d) a result \cite{schach08a}
obtained from nodal direction ARPES \cite{zhou05}.
}
\label{fig:1}
\end{figure*}

\begin{figure*}[tp]
  \begin{center}
  \includegraphics[width=8.8cm]{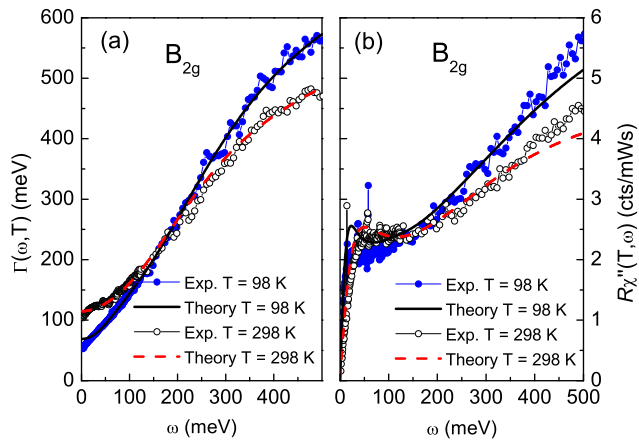}
  \end{center}
  \caption{(Colour on-line) (a) Fit to the $B_{2g}$ Raman data for the
scattering rate $\Gamma(\omega,T)$ from our maximum entropy reconstruction.
(b) The corresponding fits to the measured $B_{2g}$ cross section.
}
  \label{fig:2}
\end{figure*}

Jiang and Carbotte \cite{jiang96} give the formula for the lowest
order Raman susceptibility for an interacting electron system in the form
\begin{equation}
\label{eq:3}
  \begin{split}
     \chi_{\mu}(i\nu_n) =\\
     -T\sum\limits_m
     \sum\limits_{\bf q} Tr\left[\gamma_{\mu}^2({\bf q})\tau_3{\cal G}({\bf q},
     i\omega_m)\tau_3
     {\cal G}({\bf q},i\omega_m-i\nu_n)\right],
  \end{split}
\end{equation}
with $\omega_m$ and $\nu_n$ the fermionic and bosonic Matsubara frequencies,
respectively, \textbf{q} the momentum vector, $\tau_3$ the third Pauli
matrix and ${\cal G}({\bf q},i\omega_m)$ the electronic matrix Green's
function in Nambu notation. Equation~(\ref{eq:3}) is valid in the
superconducting as well as normal state. Here we consider only the
latter. Equation~(\ref{eq:3}) differs from the well known formula
for the optical conductivity \cite{nicol91,mars96,carb95,schach97,jiang96}
only through the factor
$\gamma^2_\mu({\bf q})$, Ref.~\cite{jiang96}, which is to be replaced
by $e^2v^2_{F,x}({\bf q})$
where $v_{F,x}({\bf q})$ is the $x$-component of the Fermi velocity at
momentum \textbf{q} which in the free electron model is assumed constant
and $v^2_{F,x}({\bf q}) = v_F^2/2$ in two dimensions. For infinite
bands with constant electronic density of states $N(0)$, Eq.~(\ref{eq:3})
can be written in a more convenient form as
\begin{equation}
  \label{eq:4}
   \begin{split}
     \chi_\mu(i\nu_n\to\nu+i0^+) =
     N(0)
     \int_{-\infty}^\infty\!d\epsilon\,\int_0^{2\pi}\!\frac{d\theta}
    {2\pi}\\
    \times\frac{[f(\epsilon)-f(\epsilon-\nu)]\gamma^2_\mu(\theta)}{\nu+
    i\tau^{-1}_{\mu,imp}+\Sigma^\star(\epsilon,\theta)-\Sigma(\epsilon+\nu,\theta)}.
  \end{split}
\end{equation}
This equation
has the same form as Eq.~(22) of Sharapov and Carbotte \cite{sharapov05}
reported in their study of the effect of the energy dependence
of the quasiparticle density of states on the far-infrared absorption
in underdoped cuprates. Equation~(\ref{eq:4}) differs as there is
now an integration over angles $\theta$, the additional
Raman vertex $\gamma^2_\mu(\theta)$, and a different numerical factor. The
same algebraic manipulation as is applied in Ref.~\cite{mars98a} gives:
\begin{eqnarray}
\label{eq:5}
\Gamma_\mu(\omega) &=& \tau^{-1}_{\mu,imp}-\frac{1}{\omega}
\int\limits_{-\infty}^\infty\!d\epsilon\,[f(\epsilon)-f(\epsilon+\omega)]
\nonumber\\
&\times&
\int\limits_0^{2\pi}\!\frac{d\theta}{2\pi}\,\gamma^2_\mu(\theta)
\textrm{Im}\left[
\Sigma_\mu(\epsilon+\omega,\theta)-\Sigma^\star_\mu(\epsilon,\theta)\right],
\end{eqnarray}
where $\Sigma_\mu(\epsilon,\theta)$ is the quasiparticle self energy due to
the directional electron-boson spectral density $I^2_\mu\chi(\Omega,\theta)$
and the $\star$ indicates the complex conjugate.
This leads directly to our fundamental Eq.~(\ref{eq:1}) when we use the
relationship for the imaginary part of the quasiparticle self energy in
terms of $I^2\chi(\Omega,\theta)$, namely
\begin{eqnarray}
  \label{eq:6}
    \textrm{Im}\Sigma_\mu(\omega,\theta) = -\frac{\pi}{2}\int\limits_0^\infty\!
    d\Omega\,I^2_\mu\chi(\Omega,\theta)\left[2\textrm{coth}\left(\frac{\Omega}{2T}
    \right)\right.\nonumber\\
    \left.-\textrm{tanh}\left(\frac{\omega+\Omega}{2T}\right)+
    \textrm{tanh}\left(\frac{\omega-\Omega}{2T}\right)\right].
\end{eqnarray}
In the approximation of Eq.~(\ref{eq:4}) for $\chi_\mu(\omega)$
we can construct the corresponding Raman cross section from
\begin{equation}
\label{eq:7}
\textrm{Im}\chi_\mu(\omega) = \chi''_\mu(\omega) =
\frac{\omega\Gamma_\mu(\omega)}
{\left\{\omega\left[1+\lambda_\mu(\omega)\right]\right\}^2+
 \Gamma_\mu^2(\omega)},
\end{equation}
where $\omega\lambda_\mu(\omega)$ is the Kramers-Kronig
transform (KK) of $\Gamma_\mu(\omega)$ as described by Opel {\it et al.}
\cite{opel00} [Eq. (A8)] to whom we refer for a detailed
discussion of how the
scattering rate $\Gamma_\mu(\omega)$ is extracted from the data
on the Raman cross section corresponding to $B_{1g}$ and $B_{2g}$ symmetries.
In Fig.~\ref{fig:1} we present our results for the electron-boson
%
%
spectral density $I^2_\mu\chi(\omega)$, Fig.~\ref{fig:1}(a) for $B_{1g}$
(antinodal) and Fig.~\ref{fig:1}(b) $B_{2g}$ (nodal) polarizations.
We show for comparison equivalent results obtained
previously from optics \cite{schach06,tu02}, Fig.~\ref{fig:1}(c), and from
nodal direction ARPES, Ref.~\cite{schach08a}, Fig.~\ref{fig:1}(d).
In Fig.~\ref{fig:2}(a) we
show our maximum entropy fits to the $B_{2g}$ Raman scattering rates for two
temperatures. The light solid (blue) dots are experiment at $T=98\,K$ and
%
%
the heavy solid (black) curve our fit. The light open (black) dots give
the data for $T=298\,$K with the heavy dashed (red) curve our fit.
Note that the low frequency part of the data at $98\,$K is almost linear,
a feature which is well captured by the theoretical
fit except for the very lowest $\omega$ region.
On the whole the general trend is well described by our data reconstruction.
To compensate for the rather big zero frequency offset of
$\Gamma_{B_{2g}}(\omega\to0,T)$ an impurity scattering rate of
$\tau^{-1}_{B_{2g},imp} = 44\,$meV had to be introduced in Eq.~(\ref{eq:1}).
In Fig.~\ref{fig:2}(b) we show a comparison between data
and theoretical fit for the $B_{2g}$ Raman cross section
(see Ref.~\cite{opel00} for notation) which
is the measured quantity from which the Raman scattering rate
$\Gamma_{B_{2g}}(T,\omega)$ of Fig.~\ref{fig:2}(a)
has been extracted. To get this
quantity we applied Eq.~(\ref{eq:7}), with $\lambda_{B_{2g}}(T,\omega)$
obtained from a Kramers-Kronig transform of our fitted Raman scattering
rate $\Gamma_{B_{2g}}(T,\omega)$. We see good agreement except for
a low frequency peak in the theoretical curve at $T=98\,$K
[heavy solid (black) curve] not present in the data. This peak is traced
to our use of an infinite band approximation. The experimental data on
$\Gamma_\mu(T,\omega)$ develop  peaks around $700\,$meV (not shown here)
for both $B_{1g}$ and $B_{2g}$ symmetry which is understood to be the result of
a reduction in the electronic density of states as a band edge \cite{schach08a}
is approached. This effect is not captured in our calculations and translates
into important differences between calculated and experimental
$\lambda_\mu(T,\omega)$. Furthermore, we observe that
theory deviates consistently from experiment to lower values for
energies greater that $\sim 350\,$meV. This can also be traced to
our use of the infinite band width approximation.
A more complete discussion of the effect of finite
bands in the inverted electron-boson spectrum can be found in
Ref.~\cite{schach08a}.

Similar results are found for the $B_{1g}$ data reconstruction.
$\Gamma_{B_{1g}}(\omega,T)$ is reproduced equally well except for some
deviations
at low energies seen in the $T=302\,$K data because of a linear
frequency dependence in the experimental $\Gamma_{B_{1g}}(\omega,T)$ data
extending from $\sim100\,$meV down to $\omega=0$ which cannot be
reproduced by theory. Such a low energy linear dependence at room
temperature has not been observed in the $B_{2g}$ data [light open (black)
dots in Fig.~\ref{fig:2}(a)].
 The zero frequency offset of $\Gamma_{B_{1g}}(\omega\to0,T)$
was compensated by an impurity scattering rate
$\tau^{-1}_{B_{1g},imp} = 81.6\,$meV
about twice as much as was necessary for $B_{2g}$. The data reconstruction
of the $B_{1g}$
Raman cross section also reveals deviations from experiment at low and high
frequencies which, again, can be understood to be the result of our infinite
band approximation on which Eq.~(\ref{eq:1}) is based.

\begin{figure*}[tp]
  \begin{center}
  \includegraphics[width=8.8cm]{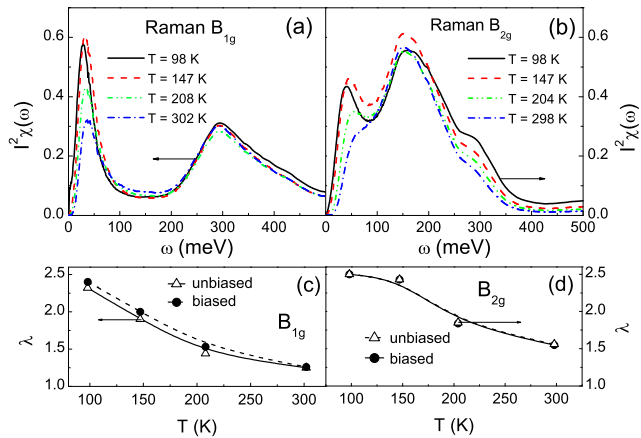}
  \end{center}
  \caption{(Colour on-line) (a) The electron-boson spectral
density $I^2\chi(\omega)$ (dimensionless) as a function of energy
$\omega$ in meV from $B_{1g}$ Raman scattering rates
using a biased maximum entropy inversion as described in the text.
(b) The same as (a) but now for $B_{2g}$ symmetry. Frames (c) and (d)
compare the temperature dependences of the mass enhancement parameters
$\lambda$ obtained from biased and unbiased inversion of the $B_{1g}$
and $B_{2g}$ Raman scattering rates, respectively.
}
  \label{fig:3}
\end{figure*}

We turn now to a comparison of antinodal with nodal Raman
results of Figs.~\ref{fig:1}(a) and \ref{fig:1}(b).
The shapes of the distributions obtained are quite
distinct. For $B_{1g}$ there is a large peak at $\sim 29\,$meV in the
$T=98\,$K spectrum followed by a dip and then a second peak around
$300\,$meV. As the temperature is increased there is a clear evolution of
the spectrum with the low energy peak decaying in amplitude, broadening
and moving towards higher energies. The same trend, although less pronounced,
is observed for the second peak. The valley between the peaks
becomes progressively filled in but it still very much remains, even
for $T=302\,$K although the effects are much less pronounced.
This is also seen when the $B_{2g}$ (nodal)
Raman spectrum is considered instead of $B_{1g}$ (antinodal). The shape
of $I^2_\mu\chi(\omega)$ for $B_{1g}$ as well as its
change with increasing temperature agrees well with previous
trends for the optical case. In Fig.~\ref{fig:1}(c)
we show results at $T=100\,$K, $200\,$K and $295\,$K in a sample of Bi2212
based on data by Tu {\it et al.} \cite{tu02}. The prominent peak is at a
position ($\sim 44\,$meV) slightly different from
our Raman spectrum, but the overall
shape at $100\,$K is in good qualitative agreement with the Raman result
[Fig.~\ref{fig:1}(a)] including the
second peak, the valley between the two peaks, and the
temperature evolution. A more extensive set of data on a similar
sample but doped with some yttrium is found in Ref.~\cite{hwang07a}.
The data presented in their Fig.~2, top frame, is also in good
qualitative agreement with our Raman results. The important
observation is that both antinodal Raman and optical data
show a prominent peak
in the electron-boson spectral density around 30 to $40\,$meV which does
not appear in ARPES and appears much less prominently in nodal Raman data.
More specifically, the fractional area under this peak is only
$\sim 3\%$ in $B_{2g}$ as compared with $\sim 23\%$ in $B_{1g}$.
As noted above, these two structures evolve with temperature in the
same way which indicates their common origin. Part of this
temperature evolution could be due to the reduced resolution intrinsic
to our unbiased inversion method as the temperature is increased.
(See Ref.~\cite{schach06}.) This is consistent with the results
obtained with the biased inversion method discussed later in
Fig.~\ref{fig:3} which shows less temperature dependence.

For nodal ARPES there
is no peak at $\sim 30\,$meV but one appears instead at much higher
energy $\sim 65\,$meV. This higher energy has often been identified
with coupling to oxygen phonons \cite{Lanzara:2001,cuk04,zhou05,schach09,
iwasawa08} and we will return to this issue later. Such a peak is not seen
in our Raman spectra which indicates that such effects are mainly
confined to the nodal direction and do not appear in averaged quantities
such as $B_{1g}$ and even $B_{2g}$ Raman even though for this latter
polarization the Raman vertex peaks in the nodal direction. The
fact that the peak at $\sim 30\,$meV is stronger in $B_{1g}$ (antinodal)
than in $B_{2g}$ (nodal) and is not seen in nodal ARPES is consistent
with a boson which is associated with scattering through momentum
transfer {\bf q} of $(\pi,\pi)$. Such a vector corresponds to transitions
between those parts on the Fermi surface which lie also on the
antiferromagnetic Brillouin zone and is, therefore, closer to the
antinodal direction. Consequently, quantities that emphasize the area
around $(\pi,0)$ such as $B_{1g}$ Raman and, to some extent, optics should
show a strong peak. In contrast, the peak at $\sim 30\,$meV  is expected
to be weaker in $B_{2g}$ symmetry weighing out the nodal part. This is
what we observe and have shown in the data of Figs.~\ref{fig:1}(a) and (b).

Instead of using a maximum entropy technique to invert Eq.~(\ref{eq:1})
van Heumen {\it et al.} \cite{heumen09} used
a histogram to characterize the electron-boson spectral density,
$I^2_{opt}\chi(\omega)$, derived from optics.
The histogram is then used to directly reconstruct the
experimental data by inversion of
Eq.~(\ref{eq:4}) modified for the optical conductivity.
They find less temperature dependence than we have here,
for the position and width of the peak around $30\,$meV in $B_{1g}$.
So far we used an {\em unbiased} maximum entropy inversion of
Eq.~(\ref{eq:1}) in which the default model \cite{schach06} is set to
a constant
at all temperatures. Another method is the so-called {\em biased}
maximum entropy inversion in which the default model is set to the
previous next lower temperature solution. We can expect
the solutions of these two methods to be different because the inversion
of Eq.~(\ref{eq:1}) is an ill-posed problem which usually has more than
one solution. It was pointed out by Yang {\it et al.} \cite{yang09} that
in the case of HgBa$_2$CuO$_{4+\delta}$ the biased inversion of optical
data resulted in electron-boson spectra very similar to those reported
by van Heumen {\it et al.} \cite{heumen09} for the same material.

Results of such a biased inversion of the Raman $B_{1g}$ and $B_{2g}$ data
are shown in Figs.~\ref{fig:3}(a) and \ref{fig:3}(b), respectively.
We see that the peak at $\sim 30\,$meV
%
%
for the $B_{1g}$ spectra shifts less with temperature than is shown
in Fig.~\ref{fig:1}(a) although it still loses
amplitude with increasing temperature $T$. This points to the
possibility that at least part of this peak
comes from a phonon, and its contribution could be determined from the
amplitude of this peak at room temperature. The situation is quite
different for the $B_{2g}$ polarization. Comparison of the spectra shown
in Fig.~\ref{fig:3}(b) with the ones presented in
Fig.~\ref{fig:1}(b) reveals that there is very little
difference between the two sets of spectra. In particular, the position
of the low energy peak
($\sim 40\,$meV at $98\,$K) shows almost the same temperature dependence
and is almost smeared out at room temperature.
Furthermore, the inversion method has little or no influence
on the shape and size of the high energy part ($\omega > 100\,$meV)
of the $B_{1g}$ and $B_{2g}$ spectra. Figures~\ref{fig:3}(c) and
\ref{fig:3}(d) show results for the often used mass enhancement parameter
$\lambda_\mu = 2\int_0^\infty\!d\omega\,I^2_\mu\chi(\omega)/\omega$
as a function of temperature for $B_{1g}$ and $B_{2g}$ polarization,
respectively. There is very little difference between the two methods
of inversion.

Returning to Fig.~\ref{fig:1}(b), the nodal direction Raman spectrum
($B_{2g}$) has a very different characteristic shape as compared to
$B_{1g}$, Fig.~\ref{fig:1}(a). While it can be characterized
as also having two peaks, the valley between them is not
pronounced and its spectral weight is much more uniformly distributed
below $300\,$meV with a relatively sharp drop off beyond this energy. This
shape is much closer to what has been found in inversions
of the nodal direction ARPES data reproduced in Fig.~\ref{fig:1}(d).
This is expected since the $B_{2g}$ Raman vertex peaks at the nodal
direction. Note that ARPES is strictly directional and samples only
the nodal direction while $B_{2g}$ Raman probes an extended part
of the BZ weighted by $\sin^2(2\theta)$. Nevertheless,
the agreement as to shape between $B_{2g}$ Raman and nodal direction ARPES
gives one confidence that both methods are measuring the same boson
spectrum. Note that there is nothing which limits the
application of this method to the high-$T_c$ cuprates and our inversion
technique has a more general applicability to other metals.

While our inversions provide us with a good handle on
the size and qualitative shape of the spectral density, the question
as to the origin of the boson involved is more difficult to answer in a
definite way. Certainly if phonons are involved we would expect
$I^2_\mu\chi(\omega)$ to mirror the phonon frequency distribution while
for spin fluctuations we should see an image of the local spin
susceptibility. The fact that $I^2_\mu\chi(\omega)$ shows very significant
spectral weight up to $400\,$meV means that excitations
other than phonons having energies below $100\,$meV
are involved. For spin fluctuations the energy scale
is set by the exchange coupling, $J$, which enters, e.g., the $t-J$ model
as a parameter, and this is consistent with the
large energy scale seen here. It is also consistent with
recent numerical studies of the $t-J$ model by Maier {\it et al.}
\cite{maier08} in which an effective electron-boson spectral density
associated with short range spin fluctuations is extracted and identified
as the pairing glue.
It displays many of the features seen in our empirical spectra. The
cellular dynamical mean-field calculations of Kyung {\it et al.} \cite{
kyung09} based on the Hubbard model also give qualitatively similar
results for the spectral density and
offer further microscopic support for an interpretation of our
derived spectra as due largely to spin fluctuations.

On the other hand one does expect, and experimental data
provide support for some contribution from the electron-phonon interaction
\cite{pasupathy08,cuk04,zhou05,meevasana06}. While
the change in critical temperature $T_c$
on substitution of $^{16}\textrm{O}\to^{18}\textrm{O}$ is small for
optimally doped samples, it
is nonzero and it can be large for the underdoped case. However, the latter
fact can also be understood as due to an energy dependence in the electronic
density of states \cite{schach90} or to a pseudogap
formation \cite{dahm00}
while at the same time the underlying contribution of phonons to
the pairing interaction remains small.  Recent
ARPES experiments along the nodal direction \cite{iwasawa08} found a shift
in the `boson kink' in the
renormalized dispersion curves of Bi2212 upon oxygen isotope substitution.
This was assigned by Schachinger {\it et al.} \cite{schach09} as a
10\% phonon contribution to the electron-boson
spectral density found in inversions \cite{schach08a} of
ARPES data as reproduced in Fig.~\ref{fig:1}(d). They
assign the peak around $65\,$meV to electron-phonon effects and the rest
due to spin fluctuations. Accounting for finite band effects
this translates into a contribution to the quasiparticle mass enhancement
due to phonons of $\lambda\simeq 0.2$ at a temperature of $T=17\,$K.
A similar estimate was found by
Devereaux {\it et al.} \cite{devereaux04} due to the buckling and breathing
phonon modes. Recent calculations based on local-density approximation
also predict mass enhancements due to phonons
\cite{heid08,savrasov96,bohnen03,giustino08} but are up to an order of
magnitude smaller than our result.

The aim of this work was to extract from the electronic
Raman cross section in $B_{1g}$ and $B_{2g}$ $(\mu)$ symmetry, the
corresponding electron-boson spectral density $I^2_\mu\chi(\omega)$.
Comparison with equivalent, previously extracted forms from optical
and nodal ARPES data shows remarkable consistency between the
results obtained with such different probes. Comparison of these results
provides information on the angular variation of the electron-boson
self energy and corresponding spectral density around the Fermi
surface. This arises because optics involves an average over the entire
Fermi surface while Raman weights predominantly the antinodal and
the nodal directions for $B_{1g}$ and $B_{2g}$ symmetry, respectively.
Only ARPES is perfectly directional.
While we now have a good handle on the size and
qualitative shape of the spectral density, the question of
what bosons might be involved is more difficult.
Certainly, the general shape of the $I^2\chi(\omega)$
should reflect the phonon frequency distribution if the electron-phonon
interaction is dominant while it should mirror the shape of the local spin
susceptibility if it is, instead, the spin fluctuations that are dominant.
The fact that all spectral functions obtained have very significant
spectral weight beyond $100\,$meV makes it clear that excitations other than
phonons are primarily involved. An energy scale set by $J$  is
more consistent with our results.

\section*{Acknowledgments}
This Research was supported in part by the Natural Sciences and Engineering
Research Council of Canada (NSERC) and by the Canadian Institute for
Advanced Research (CIFAR). The group in Garching gratefully acknowledges
  support by the German Research Foundation (DFG) via grant No. Ha~2071/3 in
  the Research Unit FOR538. E. S. acknowledges support
by the Austrian Research Fund (FWF), Vienna, contract
No. P18551-N16.
%
%

\end{document}